\documentclass[10pt]{article}

\usepackage[letterpaper]{geometry}
\usepackage{hicss}
\usepackage{times}
\usepackage{latexsym}
\usepackage{graphicx}

\usepackage{hyperref}
\usepackage{breakurl}

\usepackage{booktabs}

\title{Perceptions of Fairness and Trustworthiness Based on Explanations in Human vs. Automated Decision-Making}

\author{Jakob Schoeffer \\
  Karlsruhe Institute of Technology \\
  {\underline{jakob.schoeffer@kit.edu}} \\\And
  Yvette Machowski \\
  Karlsruhe Institute of Technology \\
  {\underline{yvette.machowski@alumni.kit.edu}}\\\And 
  Niklas Kuehl \\
  Karlsruhe Institute of Technology \\
  {\underline{niklas.kuehl@kit.edu}} \\}

\date{}

\begin{document}\sloppy

\pdfoutput=1

\maketitle

\begin{abstract}
Automated decision systems (ADS) have become ubiquitous in many high-stakes domains. Those systems typically involve sophisticated yet opaque artificial intelligence (AI) techniques that seldom allow for full comprehension of their inner workings, particularly for affected individuals. As a result, ADS are prone to deficient oversight and calibration, which can lead to undesirable (e.g., unfair) outcomes. In this work, we conduct an online study with 200 participants to examine people’s perceptions of fairness and trustworthiness towards ADS in comparison to a scenario where a human instead of an ADS makes a high-stakes decision---and we provide thorough identical explanations regarding decisions in both cases. Surprisingly, we find that people perceive ADS as fairer than human decision-makers. Our analyses also suggest that people’s AI literacy affects their perceptions, indicating that people with higher AI literacy favor ADS more strongly over human decision-makers, whereas low-AI-literacy people exhibit no significant differences in their perceptions.
\end{abstract}

\section{Introduction}
Automated decision-making has been increasingly adopted in areas such as hiring \cite{kuncel2014hiring}, lending \cite{townson2020ai}, or even policing \cite{heaven2020predictive}.
As the underlying systems, often referred to as \emph{automated decision systems} (ADS), are informing evermore high-stakes decisions, it is of utmost importance to understand their inner workings---particularly for individuals affected by their decisions, because any malfunction (e.g., unfair decisions) will have staggering consequences for them.

The reasons for adopting ADS are manifold \cite{kuncel2014hiring,harris2005automated}.
In fact, if properly designed and deployed, they are a valuable tool to combat stereotyping and thus contribute to overall social equity, e.g., in the fields of recruitment \cite{chalfin2016productivity,koivunen2019understanding}, health care \cite{grote2020ethics,triberti2020third}, or financial inclusion \cite{lepri2017tyranny}.
That said, ADS are typically based on artificial intelligence (AI)---specifically, machine learning (ML)---techniques, which leverage historical data to inform future decisions.
Now, if historical data is biased (e.g., because certain socio-demographic groups were systematically disfavored), an ADS will likely pick up and perpetuate existing patterns of unfairness \cite{feuerriegel2020fair}.
A significant body of research on algorithmic fairness from the recent past has identified such instances where ADS are causing harmful outcomes, e.g., in job ad delivery \cite{Imana21a}, facial recognition \cite{buolamwini2018gender}, recidivism prediction \cite{angwin2016machine}, or grading \cite{satariano2020british}.
These instances, among others, have likely been contributing to a recent general decline in trust towards AI \cite{Edelman2021}.

In recent years, an extensive body of research has been devoted to detecting and mitigating unfairness in ADS---mainly from a computer science viewpoint \cite{barocas2018fairness}.
A significant part of this work, however, has focused on formalizing the concept of fairness and modifying ML algorithms to satisfy different statistical equity constraints, without considering the feedback of individuals affected by automated decisions.
Among others, Srivastava et al. \cite{srivastava2019mathematical} emphasize this need for better understanding people’s attitudes towards fairness of ADS.
We argue that this research gap creates the potential for a plethora of high-impact contributions from information systems (IS) research around people’s perceptions of fairness and trustworthiness towards ADS.
This work is vital not only from a moral perspective but also regarding the effective design and implementation of ADS---with the end goal of creating decision systems that are fair, trustworthy, and, as a result, suitable for wide adoption.
To that end, we conduct a study to better understand people’s perceptions of fairness and trustworthiness towards ADS in comparison to the (hypothetical) scenario where a human instead of an ADS makes the decision.
We furthermore analyze how these perceptions may change depending on people’s background and experience with AI.

Another issue of ADS revolves around explaining automated decisions to affected individuals.
It is widely understood that opaque (i.e., black box) ML models do not allow for meaningful interpretations as to how or why certain outcomes were arrived at \cite{wanner2020much,peters2020opening}.
Prior research has also shown that explanations can be an effective tool for more transparent decision-making \cite{Meske2020,adadi2018peeking}.
Therefore, in this work, we provide study participants with thorough explanations regarding decisions---identical for both the case of the ADS and the human decision-maker.
The context of our study is lending, which is a common high-stakes application of ADS \cite{Atico2021}.

\section{Background and related work}
Harris and Davenport \cite{harris2005automated} define \emph{automated decision systems} (ADS) as systems that aim to minimize human involvement in decision-making processes.
In many cases, ADS have the potential to make more consistent decisions than humans \cite{chalfin2016productivity,koivunen2019understanding,grote2020ethics,triberti2020third,lepri2017tyranny}.
Such systems are popular in several industries, such as banking \cite{townson2020ai,harris2005automated} or hiring \cite{kuncel2014hiring,chalfin2016productivity,koivunen2019understanding,carey2016companies}---and they are emerging in new areas as well, e.g., in health care \cite{grote2020ethics,triberti2020third}.
With their increasing adoption in different high-stakes areas, it is vital to ensure that ADS reach fair and transparent decisions.
However, there have been multiple cases in the recent past where algorithms made discriminatory decisions, e.g., based on people's gender or race \cite{heaven2020predictive,buolamwini2018gender,angwin2016machine}.
Additionally, the underlying ML models are increasingly considered black boxes, making their interpretation challenging \cite{wanner2020much}.

\subsection{Explainable AI}
Despite being a popular topic of current research, \emph{explainable AI} (XAI) is a natural consequence of designing AI-based ADS and, as such, has been around at least since the 1980s \cite{lewis1982role}.
Its importance, however, keeps rising as increasingly sophisticated (and opaque) AI techniques are used to inform evermore high-stakes decisions.

Explanations can be distinguished along different dimensions.
Adadi and Berrada \cite{adadi2018peeking}, e.g., differentiate between model-specific and model-agnostic explanations.
\emph{Model-agnostic explanations} refer to methods that are not bound to a single type of ML model and are therefore more generalizable---which is why we employ them in this work.
Examples of the model-agnostic (example-based) explanation style, which provide information people can potentially act upon, are counterfactual explanations \cite{fernandez2019counterfactual}.
In brief, counterfactual explanations provide people with information regarding the minimum changes that would lead to an alternative (generally the desirable) decision.
Meske et al. \cite{Meske2020}, among others, discuss different types of explanations relevant to the IS community, particularly model-agnostic explanations.
They argue that explainability is essential for evaluating automated systems.
People affected by an automated decision may be particularly interested in explanations to assess the fairness or trustworthiness of the associated ADS.
Other popular model-agnostic explanation styles include the provision of the relevant features used by an ML model or (permutation) feature importance \cite{breiman2001random}---both of which we employ in this work since they could be plausibly provided by both human and automated decision-makers (i.e., ADS).
We refer to, e.g., Adadi and Berrada \cite{adadi2018peeking} or Goebel et al. \cite{goebel2018explainable} for more in-depth literature on the topic of XAI.

\subsection{Perceptions of fairness and trustworthiness regarding ADS}
A relatively new line of research, primarily in AI and human-computer interaction (HCI), has started focusing on perceptions of fairness and trustworthiness in automated decision-making.
Binns et al. \cite{binns2018s} and Dodge et al. \cite{dodge2019explaining}, e.g., compare fairness perceptions in ADS for distinct explanation styles.
Their works suggest differences in effectiveness of individual explanation styles---however, they also note that there does not seem to be a single best approach to explaining automated decisions.
Lee \cite{lee2018understanding} compares perceptions of fairness and trustworthiness depending on whether the decision-maker is a person or an algorithm in the context of algorithmic management, involving tasks like work scheduling or evaluation.
Their findings suggest that, among others, people perceive automated decisions as less fair and trustworthy for tasks that require typical human skills.
Lee and Baykal \cite{lee2017algorithmic} explore how algorithmic decisions are perceived in comparison to group-made decisions.
An interesting finding by Lee et al. \cite{lee2019procedural} suggests that fairness perceptions decline for some people when gaining an understanding of an algorithm if their personal fairness concepts differ from those of the algorithm.
Regarding trustworthiness, Kizilcec \cite{kizilcec2016much}, e.g., concludes that it is essential to provide the right amount of transparency for optimal trust effects.
We generally believe that this line of research would benefit significantly from novel contribution of IS scholars.

\subsection{Human vs. automated decisions}
People encounter algorithms and automation in different settings.
Therefore, it is essential to understand how this automation makes people feel and to infer the social force of algorithms.
While engineers tend to show optimism in the ability of ADS to trace and mitigate human biases and stereotypes, laypeople are often worried about AI taking over \cite{crawford2016there}.
However, Castelo et al. \cite{castelo2019task} found that in case of perceived objective decisions, people favor automated advice, and for subjective decisions they prefer human advice.
This is similar to the findings by Lee \cite{lee2018understanding}.
Yet, according to the study by Castelo et al. \cite{castelo2019task}, the perceived objectivity of a task can be altered; thus trustworthiness of and reliance on an automated decision can be increased.
Perhaps less surprisingly, Kramer et al. \cite{kramer2018people} found that people’s preference for human or AI-based decisions also depends on their prior experiences with ADS.

A major issue with ADS is that people are often unaware of their existence. Eslami et al. \cite{eslami2015always}, e.g., uncovered people’s ignorance towards the algorithm behind Facebook’s news feed.
More than half of the participants in their study were unaware of the algorithm’s manipulations, and some responded with anger and dissatisfaction.
This unawareness---apart from negative experiences \cite{buolamwini2018gender,satariano2020british}, among others---might be part of the reason why many people have such a profound aversion against algorithms \cite{Edelman2021,dietvorst2015algorithm}.
Increasing people’s awareness of ADS, e.g., by proactively disclosing the nature of the decision-maker, and considering their perceptions of these systems, may help raise acceptance in situations where ADS can make better (e.g., fairer) decisions than humans.

\subsection{Our contribution}
We aim to complement prior research (e.g., \cite{binns2018s,dodge2019explaining,lee2018understanding,lee2017algorithmic,lee2019procedural,kizilcec2016much,castelo2019task,kramer2018people}) to better understand people’s perceptions of fairness and trustworthiness towards ADS vs. human decision-makers in high-stakes settings.
Specifically, our goal is to add novel insights in the following ways: First, we integrate different model-agnostic explanations and provide them to study participants to enable them to assess the decision-making procedures.
This contrasts with most existing work, which have typically employed distinct individual explanation styles only.
Second, we provide identical model-agnostic explanations to study participants for both the case of ADS and the human decision-maker to not bias the collected responses.
Third, we examine how perceptions may change for people with high vs. low AI literacy \cite{long2020ai}.
To the best of our knowledge, the combination of the previous aspects has not been examined before.
Fourth, we consider the provider-customer context of lending, which differentiates our work from, e.g., Lee \cite{lee2018understanding}, who has analyzed the perceptions of human vs. automated decisions in algorithmic management.
Finally, we aim to bring (back) to the IS community pressing relevant issues of societal relevance, which have experienced seminal contributions mostly from other communities, such as computer science and HCI.

\section{Research hypotheses}\label{sec:hypo}
Drawing on Chan \cite{chan2011perceptions}, \emph{informational fairness} is about ``people’s expectation that they should receive adequate information on and explanation of the process and its outcomes.''
In accordance with Bélanger et al. \cite{belanger2002trustworthiness}, we define \emph{trustworthiness} as the perception of confidence in the reliability and integrity of the ADS.
People often tend to avoid algorithms and prefer a human decision-maker over an automated one, even in situations where the algorithm outperforms the person.
This phenomenon is called \emph{algorithm aversion} \cite{dietvorst2015algorithm}.
Based on this theory, as well as recent developments regarding a decline in trust towards AI \cite{Edelman2021}, we formulate our first two hypotheses, which conjecture higher perceptions of informational fairness and trustworthiness towards human decision-makers as compared to ADS:

\begin{itemize}
    \item[\textbf{H1}] People’s perceptions of informational fairness are higher when they are told the decision-maker is a human as compared to an ADS.
    \item[\textbf{H2}] People’s perceptions of trustworthiness are higher when they are told the decision-maker is a human as compared to an ADS.
\end{itemize}

Experts of a certain type of decision procedure may have different attitudes towards a decision that touches on their area of expertise than laypeople.
Wang et al. \cite{wang2020factors}, e.g., found a significant effect of general computer literacy on fairness evaluations in automated decision-making.
In the work at hand, we measure a construct that applies more directly to our context: We measure people’s \emph{AI literacy}, i.e., their ``set of competencies that enables individuals to critically evaluate AI technologies; communicate and collaborate effectively with AI; and use AI as a tool online, at home, and in the workplace'' \cite{long2020ai}.
We are interested in whether differences in people’s AI literacy change their perceptions of informational fairness and trustworthiness towards human vs. automated decision-makers.
Thus, we formulate the following additional hypotheses:

\begin{itemize}
    \item[\textbf{H3}] People’s AI literacy moderates the effect of the nature of the decision-maker (human vs. ADS) on people’s perceptions of informational fairness.
    \item[\textbf{H4}] People’s AI literacy moderates the effect of the nature of the decision-maker (human vs. ADS) on people’s perceptions of trustworthiness.
\end{itemize}

\section{Methodology}
We evaluate our hypotheses in the context of lending---an example of a provider-customer encounter.
Specifically, we confront study participants with situations where a person was denied a loan.
We argue that this is a common context that affects many people at some point in life.
According to, e.g., Atico \cite{Atico2021}, this is also an area where ADS are commonly employed for high-stakes decision-making.

\subsection{Study design}
\paragraph{Overall setup} We choose a between-subject design with the following conditions: First, we reveal to study participants some basic information about the lending company---similarly to the study setup introduced in our earlier work \cite{schoffer2021study}.
We then explain that the company rejected a given individual’s loan application.
Afterwards, we randomly allocate study participants to one of two conditions: 50\% of participants are provided the information that an ADS made the decision, and the other 50\% are told that the decision-maker was a human being.
We then provide identical explanations regarding a decision to study participants in either condition, the exact specifications of which will be derived and explained in more detail shortly.
Finally, we measure perceptions of informational fairness (INFF) and trustworthiness (TRST) through multiple measurement items, drawn (and partially adapted) from previous studies (INFF: Colquitt et al. \cite{colquitt2001justice}; TRST: Carter and Bélanger \cite{carter2005utilization}, Chiu et al. \cite{chiu2009understanding}, Lee \cite{lee2018understanding}).
Additionally, we measure AI literacy (AILIT) of study participants, with items partially derived from Long and Magerko \cite{long2020ai} as well as Wilkinson et al. \cite{wilkinson2010construction}.

\paragraph{Data and ADS}
We design and implement a functional ADS for our study---similarly to earlier work by the authors \cite{schoffer2021study}.
The ADS consists of an ML model that predicts loan approval on unseen data and can output different explanations.
For training our model, we utilize a publicly available dataset\footnote{\url{https://www.kaggle.com/altruistdelhite04/loan-prediction-problem-dataset} (last accessed: August 24, 2021)} on home loan application decisions, which has been used in multiple data science competitions on the platform \texttt{Kaggle}.\footnote{\texttt{Kaggle} is the world’s largest data science community (\url{https://www.kaggle.com/})}
The dataset at hand consists of 614 labeled (loan Y/N) observations.
It includes the following features: \emph{applicant income, co-applicant income, credit history, dependents, education, gender, loan amount, loan amount term, marital status, property area, self-employment}.
Note that comparable data---reflecting a given finance company’s circumstances and approval criteria---might, in practice, be used to train ADS \cite{Infosys2019}.
After removing data points with missing values, we are left with 480 observations, 332 of which (69.2\%) involve the positive label (Y) and 148 (30.8\%) the negative label (N).
As it is common in ML-based applications, we use 70\% of the dataset to train our ADS and use the remaining 30\% as a holdout set for the experiment.
As groundwork for the design of our ADS, after encoding and scaling the features, we train a random forest classifier \cite{breiman2001random}.
The classifier is then able to predict the (unseen) held-out labels---which it achieves with an out-of-bag accuracy of 80.1\%.
We use this classifier as a basis for the scenarios and explanations that participants are confronted with.

\paragraph{Explanations}
Recall that 50\% of study participants are assigned the \emph{ADS} condition and 50\% the \emph{human} condition.
Both conditions are provided with identical explanations regarding the decisions---the only difference is that study participants in the ADS condition are told that the ADS provides the explanatory information.
In contrast, participants in the human condition are told that a company representative (i.e., a human) provides this information.

We now explain in more detail the provided explanations.
As noted earlier, we employ only model-agnostic explanations \cite{adadi2018peeking} in a way that they could plausibly be provided by humans and ADS alike.
First, we disclose all \emph{features} (applicant income, co-applicant income, etc., as mentioned earlier), including corresponding values (e.g., \emph{applicant income: \$3,069 per month}) for an observation (i.e., an applicant) from the holdout set whom our ADS denied the loan.
We refer to such an observation as a \emph{setting}.
In our study, we employ different settings to ensure generalizability.

We also explain to study participants the \emph{importance} of these features in the decision-making process.
For that, we compute permutation feature importances \cite{breiman2001random} from our model and obtain the following hierarchy, ordered from most to least important: \emph{credit history $>$ loan amount $>$ applicant income $>$ co-applicant income $>$ property area $>$ marital status $>$ dependents $>$ education $>$ loan amount term $>$ self-employment $>$ gender}.
Note that feature importance is a global explanation style, meaning that this ordered list will be identical for any setting (i.e., applicant).

For each setting, we finally provide three \emph{counterfactual} scenarios where one actionable feature each is minimally altered such that our model predicts a loan approval instead of a rejection (e.g., \emph{the individual would have been granted the loan if, everything else unchanged, the co-applicant income had been at least \$800 per month}).
To ascertain which of the features are actionable---in a sense that people can (hypothetically) act on them to increase their chances of being granted a loan---we conducted an online survey with 20 quantitative and qualitative researchers.
According to this survey, the top-5 actionable features are \emph{loan amount, loan amount term, property area, applicant income, co-applicant income}.
We finally provide counterfactual explanations for a random subset of three of these features per setting.

\subsection{Data collection}
We conducted a between-subjects online study to test our hypotheses.
Participants for this study were recruited via \texttt{Prolific}\footnote{\texttt{Prolific} is an online platform for recruiting high-quality research participants.} \cite{palan2018prolific} and randomly assigned to either the human decision scenario or the ADS decision scenario.
Every participant was provided with two questionnaires associated with two different settings.
In each questionnaire, we asked participants to rate their agreement with multiple statements per construct on 5-point Likert scales \cite{joshi2015likert}.
A score of 1 corresponds to ``strongly disagree'' and a score of 5 to ``strongly agree''.
To be able to understand participants’ quantitative responses better, we included multiple open-ended questions as well.
We had to eliminate 4 of the 200 collected responses due to failure to pass an attention check---therefore, we analyzed 196 responses.
Among our participants, 62\% were male, 36\% female, and the remaining 2\% referred to themselves as non-binary or did not disclose their gender at all; 42\% were students, 29\% employed full-time, 11\% employed part-time, 7\% self-employed, 10\% unemployed, and 1\% chose not to disclose their profession.
The average age of participants was 26.4 years.

\section{Quantitative and qualitative results}
Before conducting our tests, we assess the validity and reliability of our latent constructs (INFF, TRST, AILIT), each of which is measured through multiple items.
We note that average variance extracted (AVE) is above or equal to the recommended threshold of 0.5 for INFF and TRST, while the AVE of AILIT is 0.39.
According to Fornell and Larcker \cite{fornell1981evaluating}, if the AVE value of a construct is low, its convergent validity can still be sufficient if composite reliability (CR) is above 0.6.
The CR of all our three constructs, INFF (0.83), TRST (0.94), and AILIT (0.72) is, in fact, above the threshold of 0.7, which is recommended by Barclay et al. \cite{barclay1995partial}.
Therefore, our convergent validity is sufficient for AILIT as well.
Values for Cronbach’s alpha (CA) are larger than the recommended threshold of 0.7 for our three constructs, indicating good reliability \cite{cortina1993coefficient}.
Validity and reliability measures are summarized in Table \ref{tab:corr_meas}.

\begin{table*}[thb]
\centering
\caption{Correlations and measurement information for latent factors. \vskip 3pt}
\label{tab:corr_meas}
\begin{tabular}{c c c c c c c c c}
\toprule
\bf Factor & \bf M & \bf SD & \bf CA & \bf CR & \bf AVE & \bf INFF & \bf TRST & \bf AILIT \\
\midrule
INFF & 3.57 & 0.62 & 0.83 & 0.83 & 0.50 & 1.00 & & \\
TRST & 3.45 & 0.72 & 0.94 & 0.94 & 0.72 & 0.69 & 1.00 &  \\
AILIT & 2.87 & 0.61 & 0.71 & 0.72 & 0.39 & 0.30 & 0.27 & 1.00 \\
\midrule
\multicolumn{9}{l}{Notes: M = Mean; SD = Standard deviation} \\
\bottomrule
\end{tabular}
\end{table*}

\subsection{Comparison of perceptions}
We conduct two Mann-Whitney U tests \cite{mcknight2010mann} to examine the differences in perceptions between ADS and human decision-makers.
The Mann-Whitney U test for informational fairness is statistically significant ($p=0.017$), suggesting a significant difference between participants’ perceptions of informational fairness.
Comparing the means of perceptions of informational fairness for both conditions reveals that the ADS condition ($M=3.68$) is perceived to be significantly fairer than the human condition ($M=3.47$).
For perceptions of trustworthiness, however, there is no significant difference between the conditions ($p=0.113$).
Hence, neither \textbf{H1} nor \textbf{H2} are supported by our analyses.
In fact, \textbf{H1} is reversely supported, eventually suggesting that for our study setup, informational fairness perceptions tend to be \emph{higher towards the ADS} compared to the human decision-maker.
Based on qualitative responses from study participants, we conjecture that this might be due to the perceived absence of emotions and subjectivity in automation.
Other potential reasons for this based on qualitative feedback are given in Section \ref{sec:qual}.
Note that this finding seems contradictory to some prior works’ results (e.g., Castelo et al. \cite{castelo2019task}), which raises doubts about the generalizability of such findings beyond specific domains.

Interestingly, when considering people’s AI literacy, these results change.
For this analysis, we split our data into two (approximately equal-sized) sub-samples along the median value of AI literacy.
We refer to one sample as \emph{high AI literacy} participants and the other as \emph{low AI literacy} participants.
We then conduct separate Mann-Whitney U tests for the two sub-samples.
Participants with high AI literacy perceive the ADS as significantly more informationally fair ($p=0.021$) and more trustworthy ($p=0.042$) than the human decision-maker.
For participants with low AI literacy, we do not find a significant difference for perceptions of informational fairness ($p=0.312$) or trustworthiness ($p=0.995$) between the human and the ADS condition.
Hence, we conclude that AI literacy has a moderating effect, which supports \textbf{H3} and \textbf{H4}.
As stated in Section \ref{sec:hypo}, we expected the moderating effect of AI literacy.
However, the finding that people with high AI literacy tend to perceive ADS as both fairer and more trustworthy than human decision-makers is not obvious to us.
On the one hand, we might think that people with high AI literacy understand such systems better and are thus less skeptical; on the other hand, it might well be the case that the same type of people are more aware of the shortcomings of ADS (e.g., \cite{heaven2020predictive,buolamwini2018gender,angwin2016machine}).

\subsection{Qualitative insights based on open-ended questions}\label{sec:qual}
We also collected unstructured textual data based on open-ended questions embedded in our questionnaires.
An in-depth analysis reveals that many study participants are convinced that automation is precisely the reason why decisions are fair (``Automated system is fair by design'').
They perceive the ADS as fair because, in their opinion, its decisions are objective: ``it [the ADS] states the criteria and follows [them], there is no room for subjectivity and the data is used to make an objective decision.''
This is likely one of the reasons why our hypotheses \textbf{H1} and \textbf{H2} are not supported.
While some participants allude to underlying issues of automated decisions (``AI can be programmed to be unfair'' and ``I do not believe an Automated Decision System can replace a human. We can't expect it to not make mistakes''), most view the ADS as fair because the system is ``purely looking at numbers [therefore] its [\emph{sic}] completely fair.''
Finally, one person points out that the situation ``is fair because the consumer knows that he has been judged using an algorithm.''

On the other hand, an interesting comment states that ``[t]he decision may have been made by a machine, but someone decided to program it that way,'' which raises questions around accountability of ADS.
Some issues are equally criticized in the human and the ADS condition: ``I don't think it is fair to take education, gender or marital status into account,'' or ``[s]ome factors are indifferent to the decision of the loan and are personal information.''
Even though overall the human condition is perceived as significantly less informationally fair than the ADS condition and people believe the ADS ``can help eliminate [...] bias,'' there are still participants who ``hope bots wont [\emph{sic}] have to decide crucial life decisions for [them].''

\section{Conclusion and outlook}
We conducted an online study with 200 participants to evaluate differences in people’s perceptions of informational fairness and trustworthiness towards human vs. automated decision-making in the high-stakes context of lending.
We provided thorough explanations to study participants, identical in both conditions (human and automated), to facilitate meaningful and unbiased responses.
Our findings suggest that within the scope of our study setup---contrary to some prior work as well as our own hypothesis---automated decisions are perceived as more informationally fair than human-made decisions.
In contrast, no significant differences were measured for trustworthiness in our case.
Based on qualitative responses, it appears that people particularly appreciate the absence of subjectivity in ADS as well as their data-driven approach.
Interestingly, our analyses also imply that people’s AI literacy affects their perceptions, given the provided explanations.
Specifically, we found that people with high AI literacy tend to perceive ADS as both fairer and more trustworthy than a human decision-maker, whereas no significant differences for either construct were detected for people with low AI literacy.

Based on our findings, we may conjecture that providing thorough explanations can enhance perceptions of fairness and trustworthiness towards ADS over human decision-makers---particularly for people with higher AI literacy.
This hypothesis will have to be tested in follow-up work.
However, we must be cognizant of the dangers of wrongful persuasion and automation biases, i.e., the tendency of people to over-rely on ADS---which might become a problem if too many (compelling) explanations about the inner workings of ADS are provided.
Future work should also account for this by examining how perceptions change when the quality of the ADS changes for the worse (e.g., by making unfair decisions) \cite{schoeffer2021appropriate}.
Other natural extensions include the consideration of domains other than lending, as well as the adoption of different explanation styles.
We hope that our work will stimulate multifaceted future research on this topic of utmost societal relevance.






\bibliographystyle{ieeetr}
\bibliography{bibliography}

\end{document}